\documentclass[aps,12pt,nofootinbib,showpacs,showkeys,preprintnumbers,amsmath,amssymb]{revtex4-1}

\usepackage{amsmath,amsfonts,amssymb,amsthm,bm,bbm,bbold,braket,color,dsfont,fancybox,hyperref,
srcltx,subfigure,ucs,yfonts,cancel,xfrac,verbatim,xcolor}
        
\usepackage{float}

\usepackage{mathtools}
\newcommand{\beq}{\begin{equation}}
\newcommand{\eeq}{\end{equation}}
\newcommand{\bea}{\begin{eqnarray}}
\newcommand{\eea}{\end{eqnarray}}
\newcommand{\beas}{\begin{eqnarray*}}
\newcommand{\eeas}{\end{eqnarray*}}
\newcommand{\bi}{\begin{itemize}}
\newcommand{\ei}{\end{itemize}}

\usepackage{amsmath,amsfonts,amssymb,amsthm,bm,bbm,bbold,braket,color,dsfont,fancybox,hyperref,
srcltx,slashed,ucs,yfonts,cancel,xfrac,verbatim,xcolor}

\usepackage[inline,shortlabels]{enumitem}
\usepackage{nicefrac}
\usepackage{multirow}

\DeclareMathAlphabet{\mathpzc}{OT1}{pzc}{m}{it}
\definecolor{gold}{rgb}{1,0.8,0}
\definecolor{nara}{rgb}{1,0.4,0.1}
\definecolor{goldo}{rgb}{1,0.7,0}
\definecolor{greeno}{rgb}{0,0.8,0}
\def\bes{\begin{subequations}}
\def\ees{\end{subequations}}
\def\be{\begin{equation}}
\def\ee{\end{equation}}
\def\bea{\begin{eqnarray}}
\def\eea{\end{eqnarray}}
\def\ba{\begin{eqnarray}}
\def\ea{\end{eqnarray}}
\def\bear{\begin{array}}
\def\eear{\end{array}}

\newcommand{\bpm}{\begin{pmatrix}}
\newcommand{\epm}{\end{pmatrix}}
\newcommand{\BM}{\left(\begin{array}}		
\newcommand{\BMC}{\left[\begin{array}}		

\newcommand{\EM}{\end{array}\right)}		
\newcommand{\EMC}{\end{array}\right]}		

	% ...

\newcommand{\com}[1]{}
\newcommand{\K}{\mathcal{K}}

\begin{document}
\begin{flushright}
\preprint{\textbf{OU-HEP-1014}}
\end{flushright}

\title{Measuring the heavy neutrino oscillations in rare W boson decays at the Large Hadron Collider} 

\author{Gorazd Cveti\v{c}$^{1}$}
\email{gorazd.cvetic@usm.cl}
\author{Arindam Das$^{2}$}
\email{arindam.das@het.phys.sci.osaka-u.ac.jp}
\author{Sebastian Tapia$^{3}$}
\email{s.tapia@cern.ch}
\author{Jilberto Zamora-Sa\'a$^{4}$}
\email{jilberto.zamora@unab.cl}

\affiliation{$^1$Department of Physics, Universidad T\'ecnica Federico Santa Mar\'ia, Valpara\'iso, Chile.}
\affiliation{$^2$Department of Physics, Osaka University, Toyonaka, Osaka 560-0043, Japan}
 \affiliation{$^3$Department of Physics, University of Illinois at Urbana-Champaign, Urbana, IL 61801, USA.}
 \affiliation{$^4$Departamento de Ciencias F\'isicas, Universidad Andres Bello,  Sazi\'e 2212, Piso 7,  Santiago, Chile.}
\begin{abstract}
Majorana neutrinos in the seesaw model can have sizable mixings through which they can be produced at the Large Hadron Collider (LHC) and show a remarkable Lepton Number Violating  (LNV) signature.
In this article we study the LNV decay of the W boson via two almost degenerate heavy on-shell Majorana neutrinos $N_j$, into three charged leptons and a light neutrino. We consider the scenario 
where the heavy neutrino masses are within $1$ GeV $\leq M_N \leq 10$ GeV. We evaluated the possibility to measure a 
LNV oscillation process in such a scenario, namely, the modulation of the quantity $d \Gamma/d L$ for the process at the LHC where
$W^{\pm} \to \mu^{\pm} N \to \mu^{\pm} \tau^{\pm} W^{\mp *}$ $ \to \mu^{\pm} \tau^{\pm} e^{\mp} \nu_e$. $L$ is the distance within the detector between the two vertices 
of the process. We found out some realistic conditions under which such a modulation could be probed at the LHC.
\end{abstract}


\maketitle

\section{Introduction}
\label{s1}
The experimental results on the neutrino oscillation phenomena \cite{Neut1, Neut2, Neut3, Neut4, Neut5, Neut6} and the flavor mixing have established the existence of the neutrino mass and flavor mixings which are the missing pieces in the Standard Model (SM).
As a result the SM needs to be extended. The seesaw extension of the Standard Model (SM) is probably the simplest idea to explain a very small neutrino mass where SM-singlet right handed heavy Majorana $(N_R^\beta)$ neutrinos induce dimension-5 
\cite{Weinberg:1979sa} operators leading to a very small light Majorana neutrino mass \cite{seesaw0,seesaw1,seesaw2,seesaw3,seesaw4,seesaw5,seesaw6}. $N_R^\beta$ couples with the SM lepton doublets $(\ell^\alpha_\beta)$ and the SM Higgs doublet $(H)$.
The relevant part of the Lagrangian is $\mathcal{L} = -y_{D}^{\alpha \beta} \overline{\ell_L^\alpha} H N_{R}^\beta - \frac{1}{2} m_N^{\alpha \beta}\overline{N_R^{C^\alpha}} N_R^\beta + \rm{H. c.}$ 
After the electroweak symmetry breaking by the vacuum expectation value, $H^T= \Big( \frac{v}{\sqrt{2}}, 0 \Big)$ the Dirac mass matrix can be obtained as $m_D=\frac{y_D v}{\sqrt{2}}$ hence the neutrino mass matrix can be written as
\begin{equation}
m_\nu=\begin{pmatrix}0&&m_D \\ m_D^T&&m_N\end{pmatrix}.
\label{mass-mat}
\end{equation}

Diagonalizing the neutrino mass matrix we get the light Majorana neutrino mass eigenvalue as $m_\nu\simeq -M_D M_N^{-1} M_D^T$. The right handed neutrinos mix with the SM light neutrinos to interact with the SM weak gauge bosons.
The variation of the seesaw scale can be possible from the intermediate scale to the electroweak scale as the Dirac Yukawa coupling $(Y_D)$ varies from the top quark Yukawa coupling $(Y_t \sim 1)$ to the scale of the electron Yukawa coupling $(Y_e \sim 10^{-6})$ \cite{Atre:2009rg,Drewes:2013gca, Deppisch:2015qwa,Cai:2017mow,Das:2018hph}. Since the heavy neutrinos are the SM-singlet candidates, they obtain the couplings with the weak gauge bosons only through the mixing via $Y_D$, making it possible to study the production of such heavy neutrinos at the collider experiments.
Historically there are a variety of search strategies of the heavy neutrinos at different existing and future facilities like the LHC, Linear Collider (LC), Large Hadron Electron Collider (LHeC)  \cite{Casas:2001sr,Das:2018usr,Das:2017nvm,Das:2012ze,Antusch:2017pkq,Antusch:2017ebe,Antusch:2016ejd,delAguila:2008cj,Cvetic:2019shl,Chakraborty:2018khw,BhupalDev:2012zg,Das:2017zjc,Das:2017rsu,Das:2014jxa,Das:2017hmg,Dev:2013wba,Boiarska:2019jcw,Bondarenko:2019tss}, tau and meson factories where the bounds on the heavy neutrino mass and the mixing with the light species have been shown \cite{Cvetic:2013eza,Cvetic:2014nla,Cvetic:2015naa,Zamora-Saa:2016qlk,Zamora-Saa:2016ito,Dib:2019tuj,Zamora-Saa:2019naq,Kim:2017pra,Mandal:2017tab,Milanes:2018aku,Abada:2017jjx,Mejia-Guisao:2017gqp}. Due to the Majorana nature of the heavy neutrinos we can obtain a pair of like sign leptons in the final state where the one lepton is produced in association with the heavy neutrino and the other one comes from the leading decay mode of the heavy neutrino into a lepton and a $W$ boson. Such a LNV signature is very distinctive for the heavy neutrinos. In addition to that, CP violation in the neutrino sector is also a crucial point for the leptogenesis, see  \cite{Chun:2017spz} for review.

In a previous article \cite{Cvetic:2018elt} we have studied the W boson decay into an LNV channel via two almost degenerate heavy on-shell Majorana neutrinos which oscillate among themselves (c.f. \cite{Blasone:1995zc,Blasone:1998hf,Naumov:2009zza,Naumov:2010um,Boyanovsky:2014una,Cvetic:2015ura,Anamiati:2016uxp}). The final state consisted of three charged leptons and a light neutrino. The third lepton and the neutrino are coming from the leptonic decay of the $W$ boson produced from the leading decay mode of the heavy neutrino $(N \to W\ell)$. We have found that due to a small mass difference $(\Delta M_N \sim \Gamma_N)$ between the heavy neutrino states the oscillation effects can be present in the decay. In the current article we focus on the scenario with at least two heavy neutrinos with $M_N \leq 15$ GeV to study in detail the effect of the oscillation. We study the effect in the LHC environment considering the quarks in the initial states. 

The article is arranged in the following way. In Sec.~\ref{s3} we study the production of the heavy neutrino in the LHC. In Sec.~\ref{sec:simu} we simulate the events of the heavy neutrino production at the LHC to study the kinematical parameters. In Sec.~\ref{sec:dis} we discuss the results and present conclusions.
\section{Production of the RHN}
\label{s3}
\begin{figure}
\centering
\includegraphics[scale = 0.7]{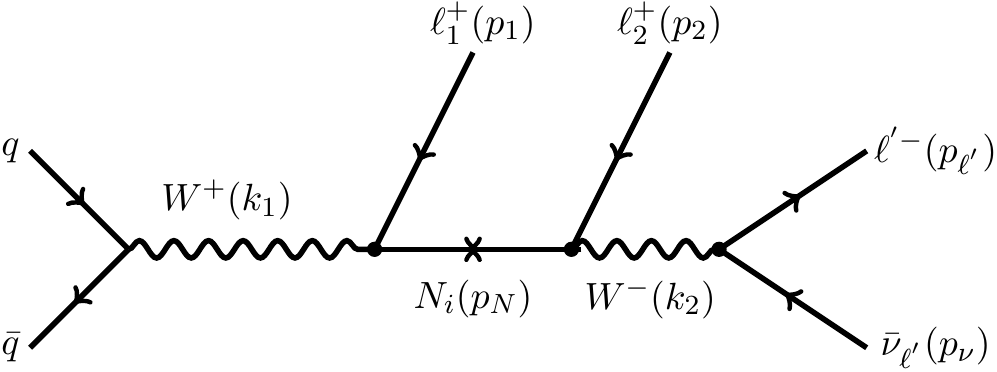}\hspace{0.3 cm}
\includegraphics[scale = 0.7]{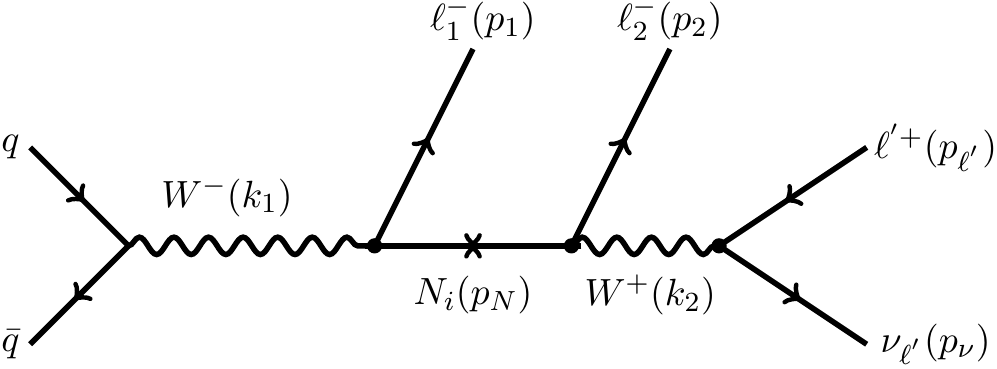}
\caption{Heavy neutrino production in the charged Drell-Yan channel. Left Panel: Feynman diagrams for the LNV process $W^+\rightarrow \ell_1^+ \ell_2^+ \ell^{' -} \bar{\nu}_{\ell^{'}}$. Right Panel: Feynman diagrams for the LNV process $W^-\rightarrow \ell_1^- \ell_2^- \ell^{' +} \nu_{\ell^{'}}$}
\label{fig:Wdecays}
\end{figure}

\begin{figure}
\centering
\includegraphics[scale = 0.65]{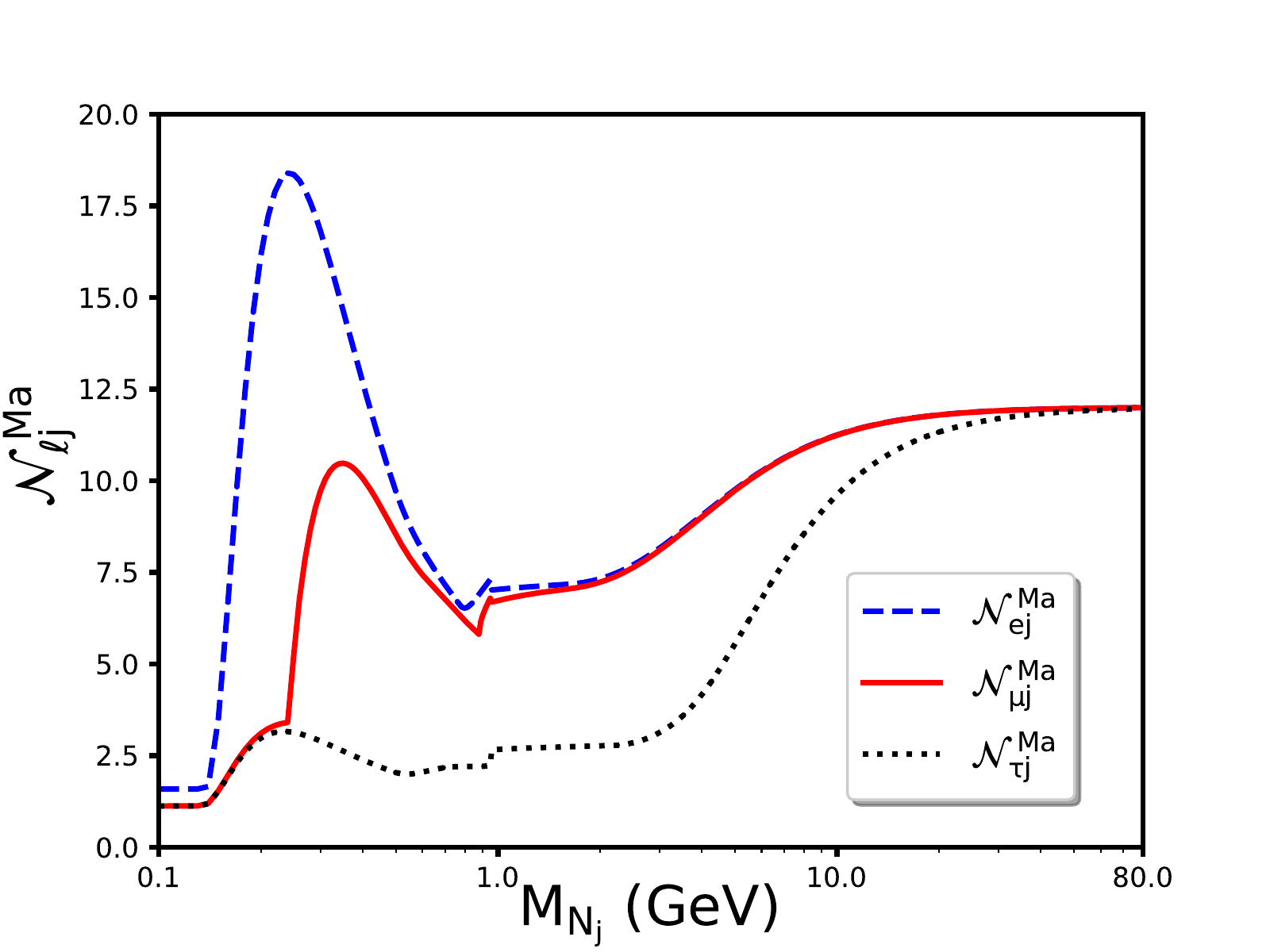}
\caption{Effective mixing coefficients ${\cal N}_{\ell j}^{\rm Ma}$ for Majorana neutrinos. Figure taken from \cite{Zamora-Saa:2016ito}.}
\label{fig:efcoef}
\end{figure}

As we stated in our previous article \cite{Cvetic:2018elt}, we are interested in studying the LNV processes which are described by the Feynman diagrams in Fig.~\ref{fig:Wdecays}. From now on, we will consider the case when $\ell_1 = \mu$, $\ell_2 = \tau$ and $\ell^{'} = e$ ($m_e \approx 0$), the heavy neutrinos $N_1$ and $N_2$ are almost degenerate and  mass difference ($|\Delta M_N|=M_{N2}-M_{N1}$) is in the range $\Gamma_{N} \leq |\Delta M_N| \leq 15~\Gamma_{N}$.
The relevant equations for such processes were presented in \cite{Cvetic:2015ura,Cvetic:2018elt} and the obtained $L$-dependent effective differential decay width considering heavy neutrinos oscillations was \cite{Cvetic:2018elt}\footnote{In Ref.~\cite{Cvetic:2015ura} this expression was obtained in the approximation when $L \ll (\gamma_N \beta_N)/\Gamma_{\rm Ma}(M_N)$.}
\begin{small}
\begin{align}
\nonumber \frac{d}{dL}\;& \Gamma(W^{\pm})  = \frac{ 1}{\gamma_N \ \beta_N} \exp\Big[-\frac{L \ \Gamma_{\rm Ma}(M_N)}{\gamma_N \ \beta_N}\Big] \; \widetilde{\Gamma}\big( W^+ \to \ell^+_1 N \big) \ \widetilde{\Gamma}\big( N \to \ell^+_2 e^- \bar{\nu}  \big)  \\
& \times \Bigg( \sum_{i=1}^2 |B_{\mu N_i}|^2 |B_{\tau N_i}|^2+ 2 |B_{\mu N_1}| |B_{\tau N_1}| |B_{\mu N_2}| |B_{\tau N_2}| \cos \Big(2\pi \; \frac{L}{L_{\rm osc}} \pm \theta_{LV} \Big)\Bigg) \ .
\label{effdwfosc}
\end{align}
\end{small}
Here, the angle $\theta_{LV}$ stands for the CP-violating phase, $\Gamma_{N} = (1/2)(\Gamma_{N_1} + \Gamma_{N_2})$ is the (average of the) total decay width of the intermediate Majorana neutrino and $L_{\rm osc} = (2 \pi \beta_N \gamma_N)/\Delta M_N$ with $\Delta M_N=M_{N2}-M_{N1}\equiv Y \Gamma_{N}$, where $Y$ is a parameter which measures the mass difference in terms of $\Gamma_N$. Then $ \Gamma_{\rm Ma}(M_{N_i})$ can be written as 
\begin{equation}
  \Gamma_{\rm Ma}(M_{N_i}) \equiv \Gamma_{N_i}   \approx  \K_i^{\rm Ma}\ \frac{G_F^2 M_{N_i}^5}{96\pi^3} 
\label{DNwidth}
\end{equation}
 with 
 \begin{equation} 
 \K_i^{\rm Ma} = {\cal N}_{e i}^{\rm Ma} \; |B_{e N_i}|^2 + {\cal N}_{\mu i}^{\rm Ma} \; |B_{\mu N_i}|^2 + {\cal N}_{\tau i}^{\rm Ma} \; |B_{\tau N_i}|^2,
\label{DNwidth1}
\end{equation}
where ${\cal N}_{\ell i}^{\rm Ma}$ are the effective mixing coefficients which account for all possible decay channels of $N_i$ and are presented in Fig.~\ref{fig:efcoef} for our mass range of interest (${\cal N}_{\ell i}^{\rm Ma} \sim 1$-$10$).
We notice that, the mixings\footnote{In some literature the mixing factors $B_{\ell N}$ are also defined as $U_{\ell N}$ or $V_{\ell N}$  (i.e. $B_{\ell N} \equiv  U_{\ell N} \equiv  V_{\ell N}$). } $B_{\ell N_1}$ and $B_{\ell N_2}$ can be, in principle, different for the two neutrinos, and therefore the two mixing factors $\mathcal{K}^{\rm Ma}_i$ $(i = 1, 2)$ may differ significantly from each other. However, from now on we will assume that  $\mathcal{K}^{\rm Ma}_1 \approx \mathcal{K}^{\rm Ma}_2$ $(\equiv \mathcal{K})$. For the heavy neutrino mass range study in this work  we will assume $|B_{\mu N_i}|^2 \approx |B_{\tau N_i}|^2 = 10^{-6}$ ($\equiv |B_{\ell N}|^2$), $|B_{e N_i}|^2=0$,  and ${\cal N}_{\mu i}^{\rm Ma} + {\cal N}_{\tau i}^{\rm Ma} \approx 20$; therefore, $\mathcal{K} = 20~|B_{\ell N}|^2$. Therefore, both heavy neutrinos are considered to have the same total decay width (note also that their masses are almost equal, $M_{N_i}=M_N$)
\begin{equation}
\Gamma_{\rm Ma}(M_{N_i}) \equiv \Gamma_N(M_N) = 20 |B_{\ell N}|^2 \ \frac{G_F^2 M_{N}^5}{96\pi^3} \, ,
\label{DNwidthappr}
\end{equation}
In Ref.~\cite{Cvetic:2018elt} we considered that the kinematical parameters (velocity $\beta_N$ and Lorentz factor $\gamma_N$) of the produced $N_j$'s in the laboratory frame ($\Sigma$) are fixed, usually $\beta_N \gamma_N=2$. However, the product $\beta_N \gamma_N$ is in general not fixed, because $W$ is moving in the lab frame when it decays into $N$ and $\ell_1$. This factor is then written as
\begin{equation}
\beta_N \gamma_N = \sqrt{(E_N({\hat p}'_N)/M_N)^2 - 1},
\label{bNgN}
\end{equation}
where the energy of neutrino $N$ in the lab frame, $E_N$, depends on its direction ${\hat p}'_N$ in the $W$-rest frame ($\Sigma'$).
\begin{figure}
\centering
\includegraphics[width=8cm]{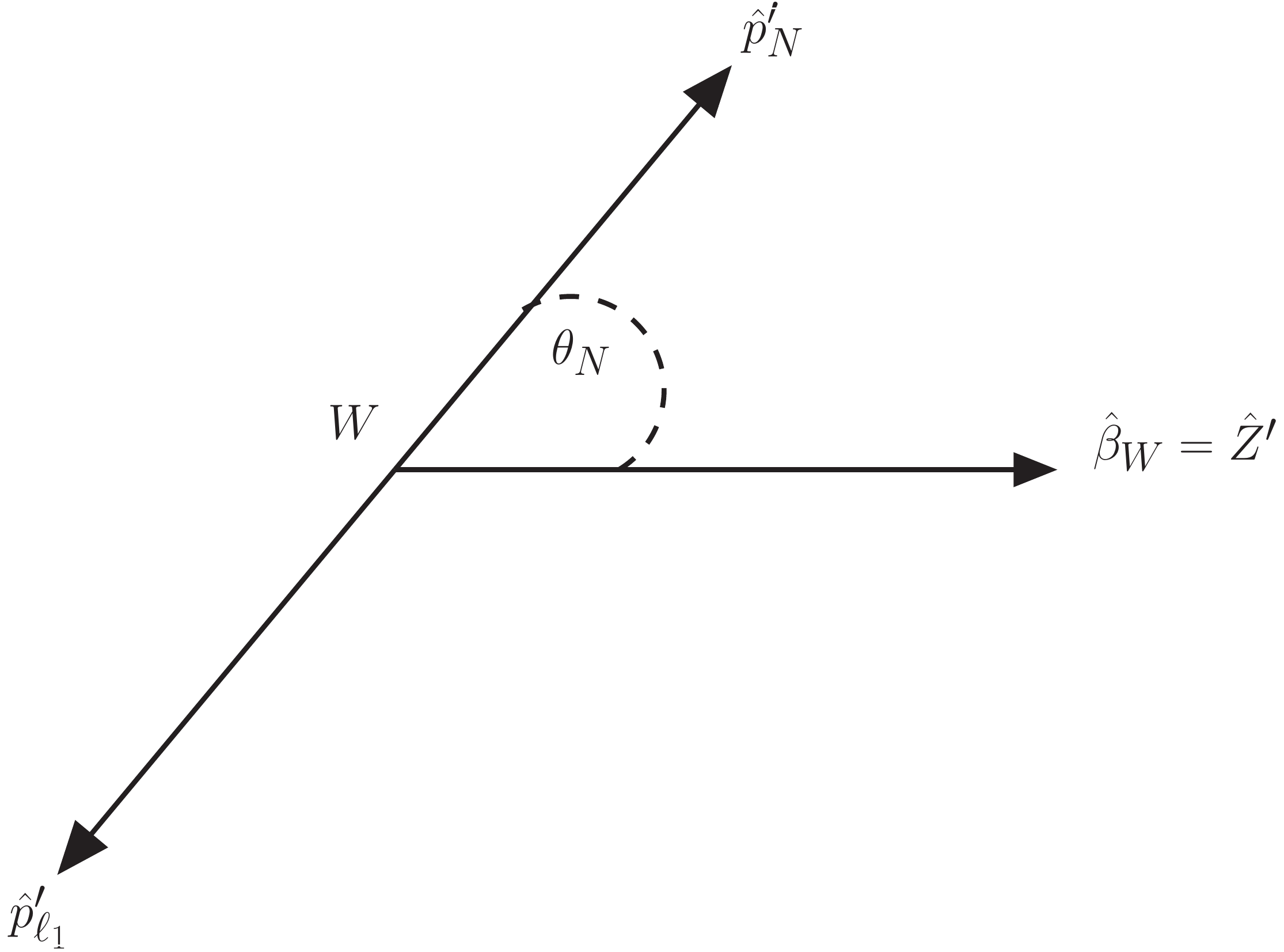}
\caption{The direction of the momentum of the produced $N$ in the $W$-rest frame ($\Sigma'$), ${\hat p}'_N$. The direction ${\hat \beta}_W$ of the velocity of $W$ in the lab frame defines the ${\hat z}'$-axis; $\theta_N$ is the angle between ${\hat \beta}_W$ and ${\hat p}'_N$.}
 \label{FigpN}
\end{figure}
The relation between $E_N$ and the angle $\theta_N$ ($\leftrightarrow {\hat p}'_N$, cf.~Fig.~\ref{FigpN}) is
\begin{equation}
E_N = \gamma_W (E'_N + \cos \theta_N \beta_W |{\vec p}'_N|),
\label{EN}
\end{equation}
where the corresponding quantities in the $W$-rest frame ($\Sigma'$) are fixed
\begin{equation}
E'_N = \frac{M_W^2 + M_N^2 - M_{\ell_1}^2}{2 M_W}, \quad
|{\vec p}'_N| = \frac{1}{2} M_W \lambda^{1/2} \left( 1, \frac{M_{\ell_1}^2}{M_W^2}, \frac{M_N^2}{M_W^2} \right),
\label{ENppNp}
\end{equation}
$\beta_W$ is the velocity of $W$ in the lab frame, and $\lambda$ is the usual phase space function.
Therefore, the formulas for the oscillation decay widths of heavy neutrinos must be written in differential form and integrated over the directions of heavy neutrino in the $W$-rest frame \cite{Cvetic:2018elt}
\begin{align}
\frac{d}{dL}\Gamma_{LV}^{\rm osc}(W^{\pm})=&\int d \Omega_{{\hat p}'_N} \frac{1}{\left[ (E_N({\hat p}'_N)/M_N)^2 - 1 \right]^{1/2} }  \exp\Big[-\frac{L \ \Gamma_N(M_N)}{\left[ (E_N({\hat p}'_N)/M_N)^2 - 1 \right]^{1/2}}\Big]  \nonumber \\
& \frac{d \widetilde{\Gamma}\big( W^+ \to \ell^+_1 N \big)}{ d \Omega_{{\hat p}'_N}} \ \widetilde{\Gamma}\big( N \to \ell^+_2 e^- \bar{\nu}  \big) \nonumber \\
& \times \Bigg( \sum_{i=1}^2 |B_{\mu N_i}|^2 |B_{\tau N_i}|^2+ 2 |B_{\mu N_1}| |B_{\tau N_1}| |B_{\mu N_2}| |B_{\tau N_2}| \cos \Big(2\pi \; \frac{L}{L_{\rm osc}({\hat p}'_N)} \pm \theta_{LV} \Big)\Bigg), 
\label{effdwfosc2}
\end{align}
where now the oscillation length $L_{\rm osc}$, appearing in the last term, also depends on the direction ${\hat p}'_N$
\begin{equation}
L_{\rm osc}({\hat p}'_N) = \frac{2 \pi\beta_N \gamma_N}{M_N} = 
\frac{2 \pi}{M_N^2} |{\vec p}_N({\hat p}'_N)| =
\frac{2 \pi}{M_N} \left[ (E_N({\hat p}'_N)/M_N)^2 - 1 \right]^{1/2}.
\label{Losc2}
\end{equation}
When we integrate the expression in Eq.~\ref{effdwfosc2} over $dL$ up to the length $L$, we obtain
\begin{small}
\begin{align}
\label{oscWfulldw}
\nonumber \Gamma_{LV}^{{\rm osc}}(W^{\pm}) =  &
\int d \Omega_{{\hat p}'_N}
\frac{d \widetilde{\Gamma}\big( W^+ \to \mu^+ N \big)}{ d \Omega_{{\hat p}'_N} }  \frac{\widetilde{\Gamma}\big( N \to \tau^+ e^- \bar{\nu}_e  \big)}{\Gamma_N(M_{N})}  {\Bigg \{}
\left( 
1 - \exp \left[ - \frac{L \Gamma_N(M_N)}{ \left[ (E_N({\hat p}'_N)/M_N)^2 - 1 \right]^{1/2} } \right] 
\right)
\times
\nonumber\\ &
\sum_{i=1}^{2} |B_{\mu N_i}|^2 |B_{\tau N_i}|^2
+ \frac{2}{y^2} |B_{\mu N_1}| |B_{\tau N_1}|  |B_{\mu N_2}| |B_{\tau N_2}|
\nonumber\\ &
{\Bigg (}  
\cos (\theta_{LV})  \mp  y \sin (\theta_{LV}) 
+ \exp \left[ - \frac{L \Gamma_N(M_N)}{\left[ (E_N({\hat p}'_N)/M_N)^2 - 1 \right]^{1/2}} \right] \times
\nonumber\\ &
\left[
y \sin \left( 2 \pi \frac{L}{L_{\rm osc}({\hat p}'_N)} \pm \theta_{LV} \right) -
\cos\left( 2 \pi \frac{L}{L_{\rm osc}({\hat p}'_N)} \pm \theta_{LV} \right) 
\right]
{\Bigg )}
{\Bigg \}}.
\end{align}
\end{small}
In Eq.~\ref{oscWfulldw}, the relative corrections ${\cal O}(1/y^2)$ were neglected. The expressions in Eqs.~\ref{effdwfosc2} and \ref{oscWfulldw} get somewhat simplified when, in the differential decay width factor $d {\widetilde{\Gamma}} ( W^+ \to \mu^+ N )/d \Omega_{{\hat p}'_N}$, we perform average over the initial polarizations of $W$ and sum over the helicities of $\mu^+$ and $N$. Namely, in such a case this differential decay width is constant (independent of the direction ${\hat p}'_N$)
\bes
\label{dGWmuN}
\bea
\frac{d {\widetilde{\Gamma}} ( W^+ \to \mu^+ N )}{d \Omega_{{\hat p}'_N}} &=&
\frac{1}{4 \pi} {\widetilde{\Gamma}} ( W^+ \to \mu^+ N )
\\
& = & \frac{1}{48 \pi^2} \frac{G_F M_W^3}{\sqrt{2}} \left[ 2 - (x_N + x_{\mu}) - (x_N-x_{\mu})^2 \right] \lambda^{1/2}(1,x_N,x_{\mu}),
\eea
\ees
where $x_N = M_N^2/M_W^2$ and $x_{\mu}=m_{\mu}^2/M_W^2$. The dependence on the direction ${\hat p}'_N$ in the expressions in Eqs.~\ref{effdwfosc2} and \ref{oscWfulldw} is then only the dependence on $\theta_N$, (cf.~Eqs.~\ref{EN} and \ref{ENppNp}); the integration $d \Omega_{{\hat p}'_N}$ then reduces to $2 \pi d \cos \theta_N$.

Furthermore, for the correct evaluation of the quantities in Eq.~\ref{effdwfosc2} and \ref{oscWfulldw}, we need to make a weighted average over various velocities $\beta_W$ (or 3-momenta $|{\vec p}_W|$) of the on-shell $W$ in the lab frame; i.e., we need information about the $|{\vec p}_W|$-distribution of the produced $W$'s in the lab frame of LHC. On the other hand, we use the expression of the decay width ${\widetilde{\Gamma}}( N \to \tau^+ e^- \bar{\nu}_e )$ from  Ref. \cite{Cvetic:2018elt}, i.e., the expression in which the structure of the off-shell $W^{-*}$ propagator is accounted in its full  form (but not in the effective form), because in the considered cases the mass $M_N$  can be comparable to the mass $M_W$.

On the other hand, the CP violating phase ($\theta_{LV}$) can be extracted by means of the difference between the $L$-dependent effective differential decay width for $W^+$ and $W^-$
\begin{small}
\begin{align}
\nonumber
\frac{d}{dL} \Gamma(W^{+})-\frac{d}{dL} \Gamma(W^{-}) =& \frac{ -1}{\gamma_N \ \beta_N} \exp\Big[-\frac{L \ \Gamma_{\rm Ma}(M_N)}{\gamma_N \ \beta_N}\Big] \; \widetilde{\Gamma}\big( W^+ \to \ell^+_1 N \big) \ \widetilde{\Gamma}\big( N \to \ell^+_2 e^- \bar{\nu}  \big)  \\
& \times 4 |B_{\mu N_1}| |B_{\tau N_1}| |B_{\mu N_2}| |B_{\tau N_2}| \sin \Big(2\pi \; \frac{L}{L_{\rm osc}} \Big) \sin \Big(\theta_{LV} \Big)  \ ,
\label{effassym}
\end{align}
\end{small}
where we used, for simplicity, the schematic formula Eq.~\ref{effdwfosc} instead of  Eq.~\ref{effdwfosc2}. In addition,
in Eq.~\eqref{effassym} it was assumed (approximated) that $\gamma_N \beta_N$ is the same for processes involving $W^+$ and $W^-$. 
\section{Heavy neutrino simulations and results}
\label{sec:simu}
\begin{figure}
\centering
\includegraphics[scale = 0.5]{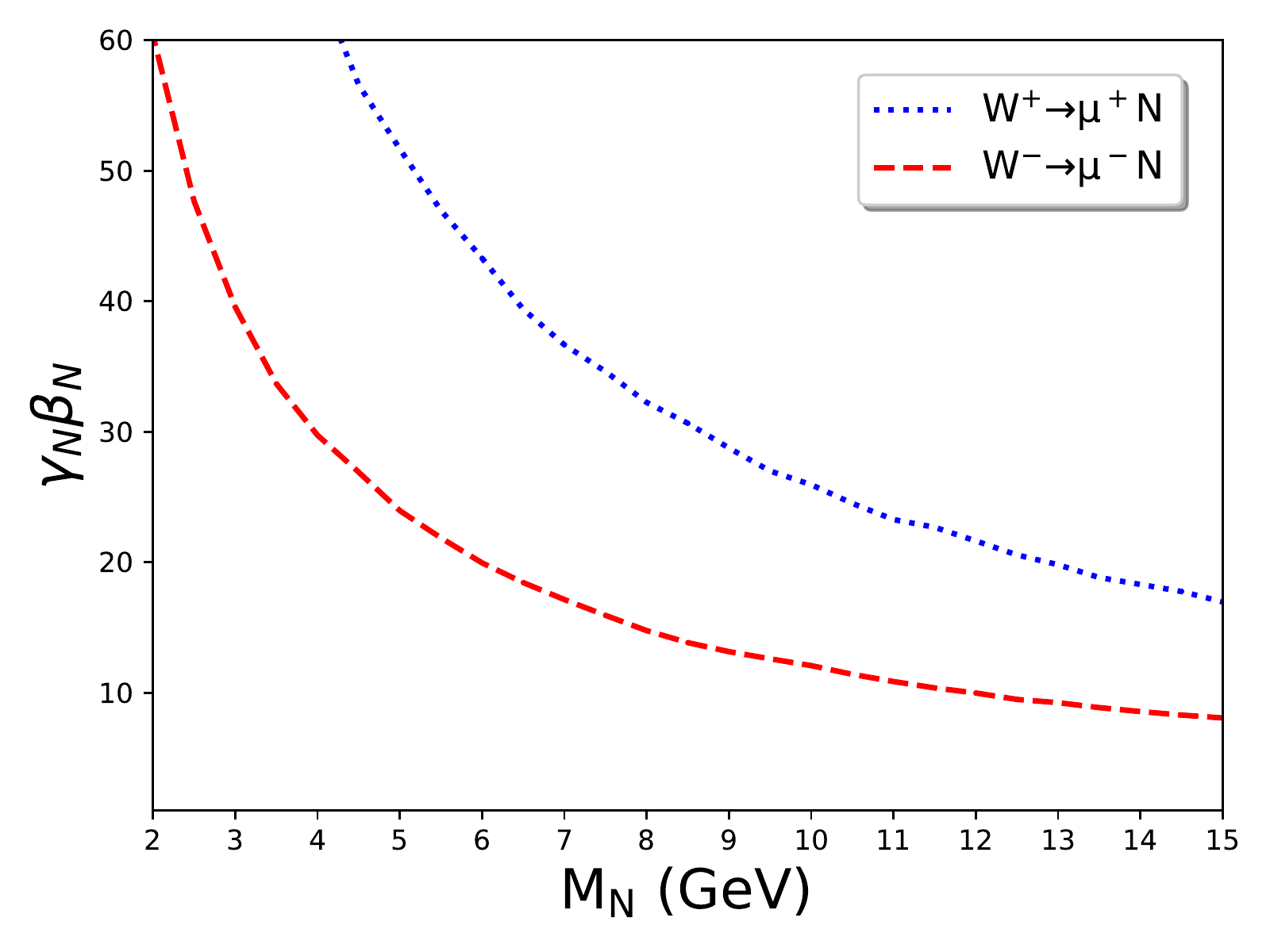}\hspace{0.01 cm}
\includegraphics[scale = 0.5]{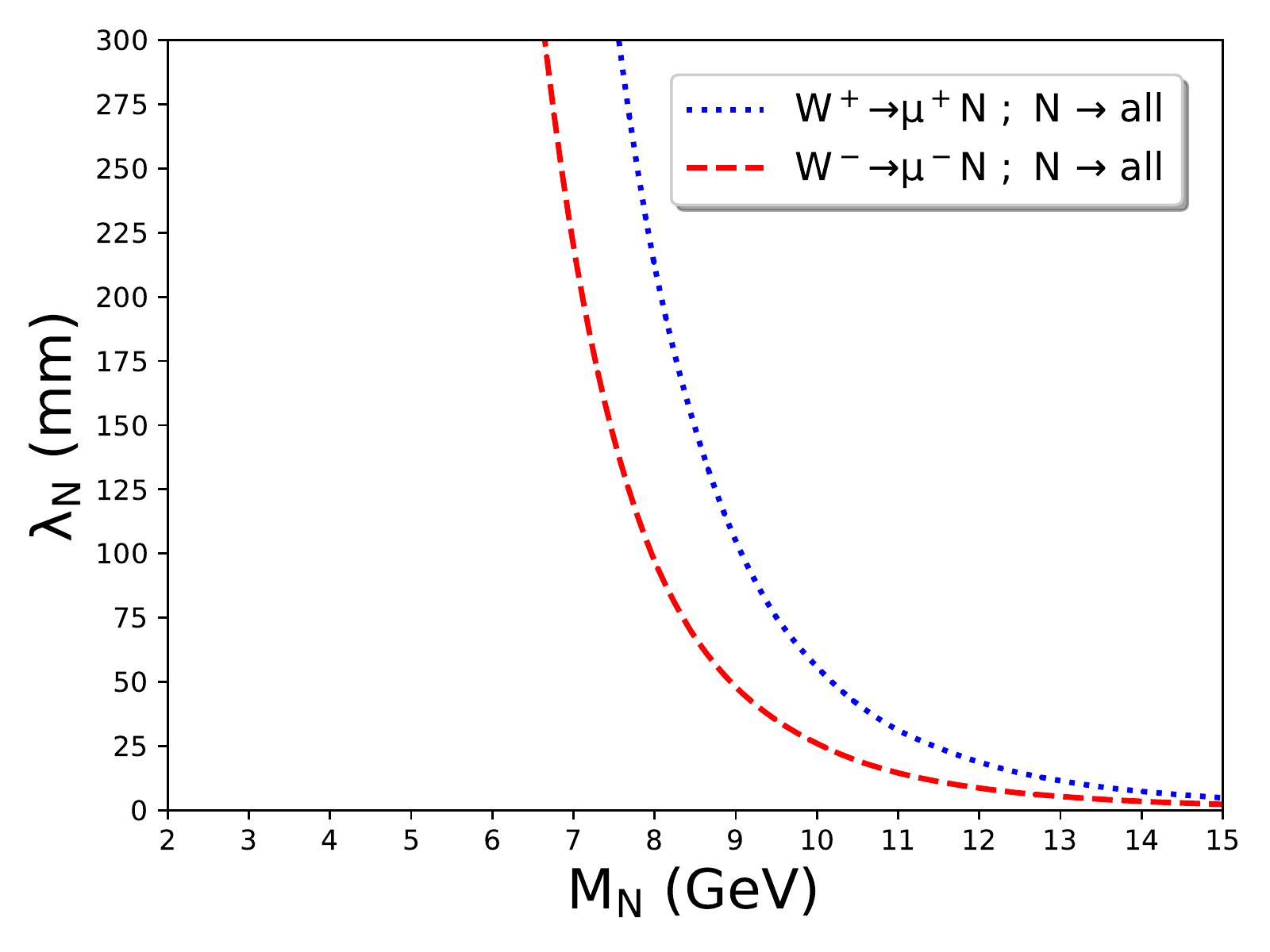}
\caption{{Left Panel}: Average heavy neutrino $\gamma_N \beta_N (= |{\vec p}_N|/M_N$) factor. Here we have taken $|B_{\ell N}|=10^{-6}$. {Right Panel}: 
Average heavy neutrino decay length $\lambda_N (= \gamma_N \beta_N/\Gamma_N$). }
\label{fig:lambda}
\end{figure}

In order to test the feasibility to measure the heavy neutrino oscillation, described by Eq.~\ref{effdwfosc2}, we need to get the correct distribution 
of $\gamma_{N}\beta_{N} = |{\vec p}_N|/M_N$ in which the heavy neutrino was produced. To obtain a realistic distribution of the $\gamma_{N}\beta_{N}$ factor, 
we simulate the heavy neutrino production via charged current Drell-Yan process shown in Fig.~\ref{fig:Wdecays}, using 
 \textsc{MadGraph5\_aMC@NLO}~\cite{Alwall:2014hca} for $W^{+}$ and $W^{-}$ individually, 
 for LHC with $\sqrt{s}=13$ TeV. The $W^+$ and $W^-$ show a significant difference in the distribution 
 of $\gamma_{N}\beta_{N}$ which has been shown in the left panel of Fig.~\ref{fig:lambda}. The Universal FeynRules Output (UFO)~\cite{Degrande:2011ua} files were generated
using the FeynRules libraries~\cite{Alloul:2013bka}. The simulation accounts for the distribution of $\beta_W$ and $\theta_N$ (cf.~Eqs.~\ref{bNgN} and \ref{EN}).

\begin{figure}
\centering
\includegraphics[width=0.49\textwidth]{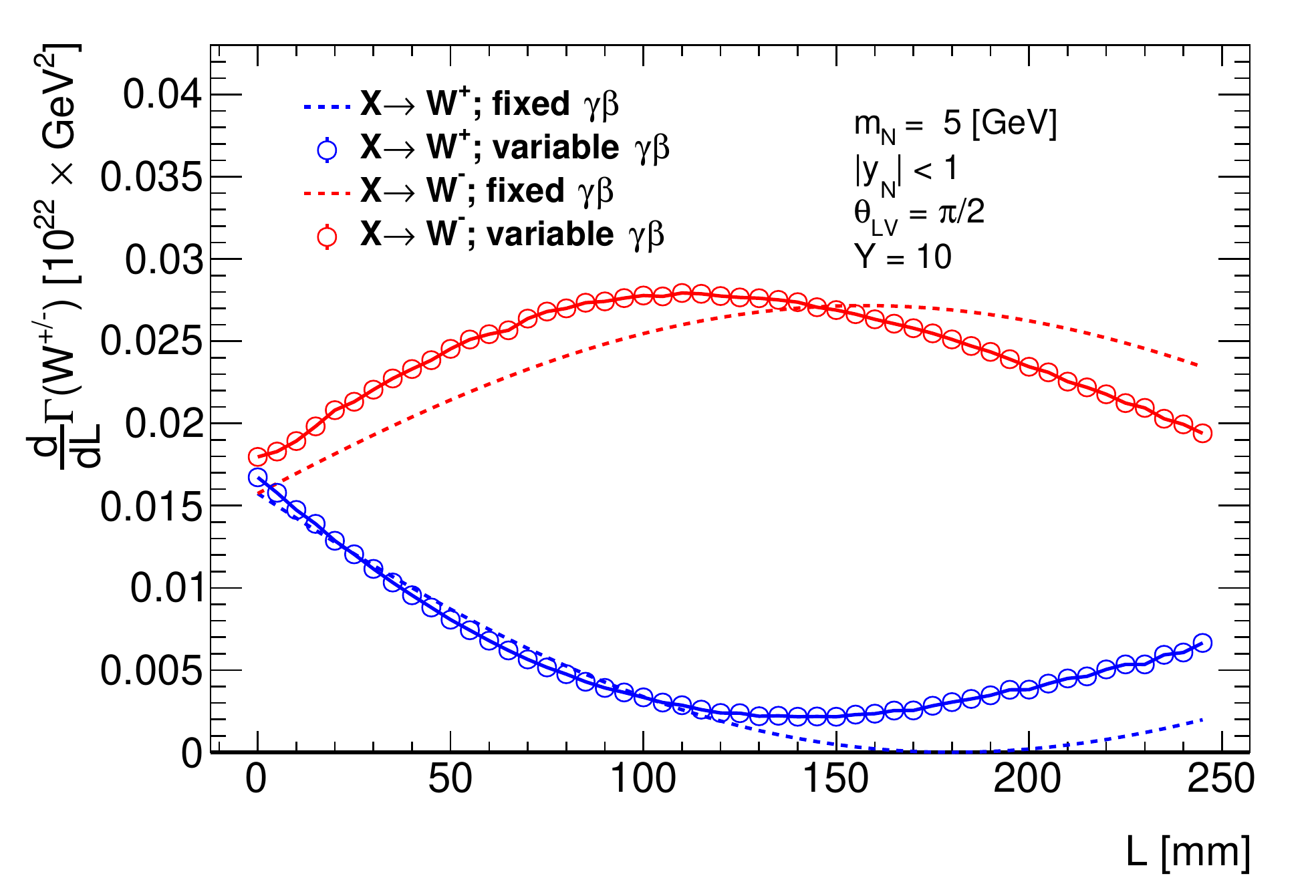}\hspace{0.01 cm}
\includegraphics[width=0.49\textwidth]{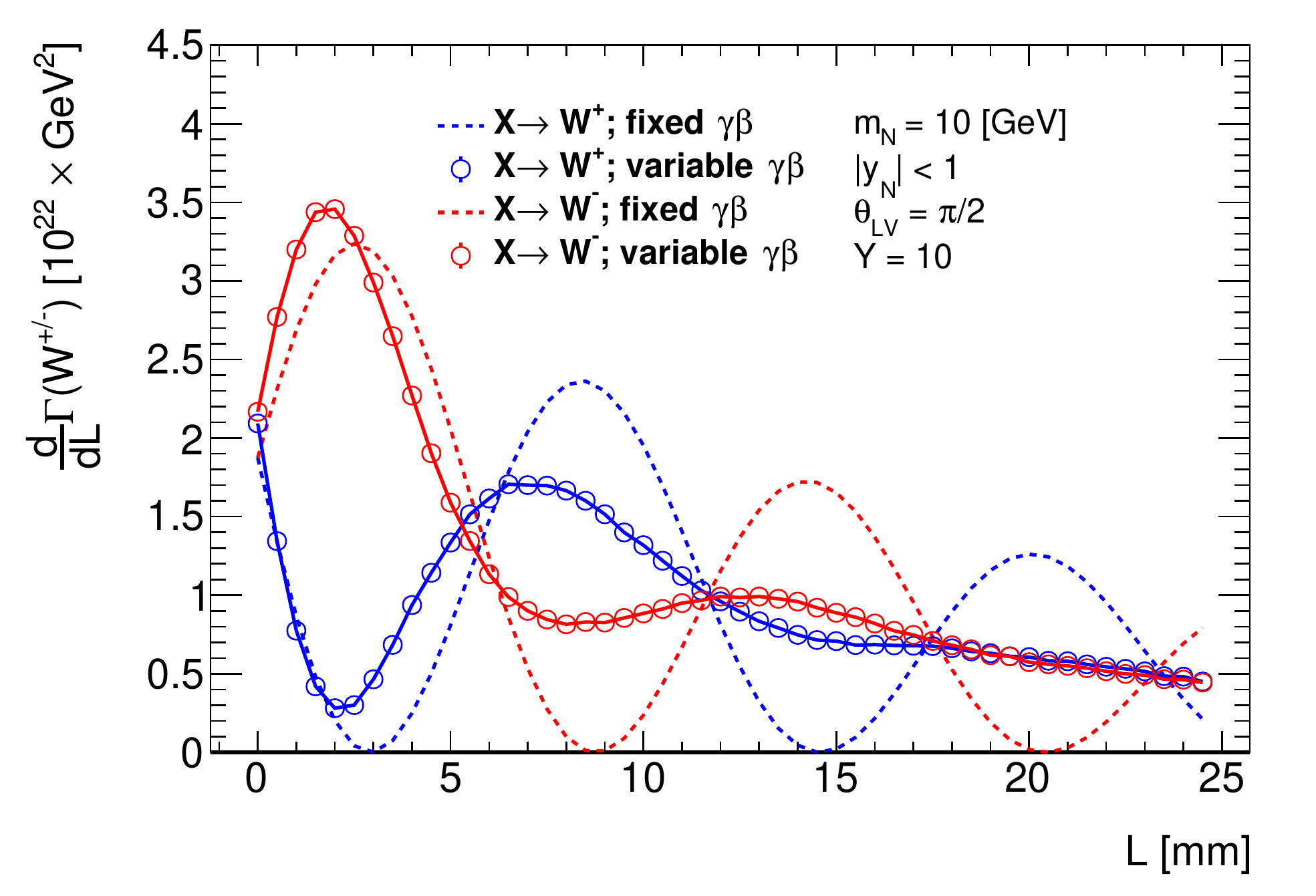}
\caption{Comparison of $d\Gamma/dL$ when we use a fixed value of $\gamma_{N}\beta_{N}$, and $d\Gamma/dL$ re-evaluated using the random sampling 
of $\gamma_{N}\beta_{N}$ from the simulation. Left panel: $M_N=5$ GeV, $\theta_{LV}=\pi/2$, $Y(\equiv \Delta M_N/\Gamma_N)=10$ and $|y_{N}| < 1$. Right panel: $M_N=10$ GeV, 
$\theta_{LV}=\pi/2$, $Y=10$ and $|y_{N}| < 1$}
\label{fig:fixvarcom}
\end{figure}

\begin{figure}
\centering
\includegraphics[width=0.49\textwidth]{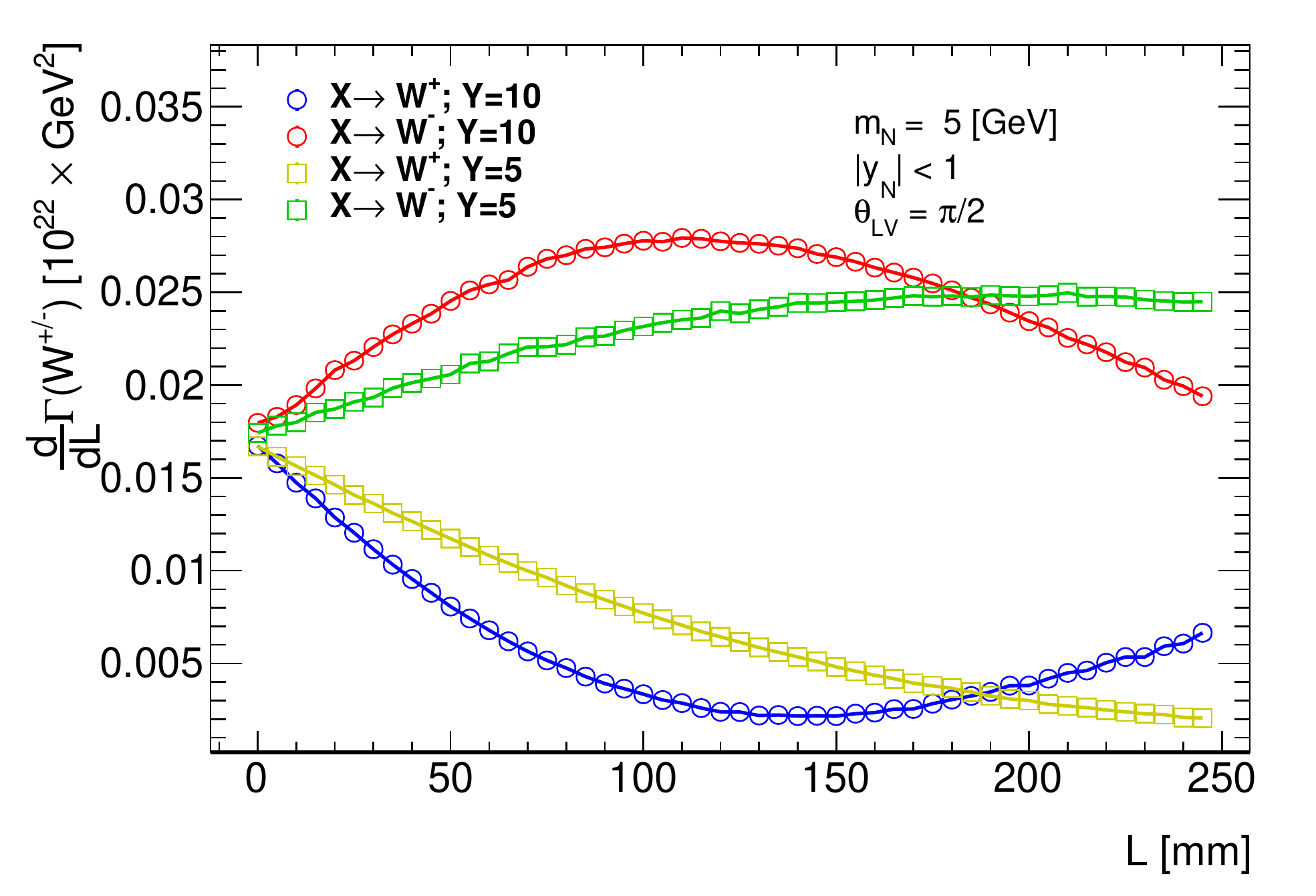}\hspace{0.01 cm}
\includegraphics[width=0.49\textwidth]{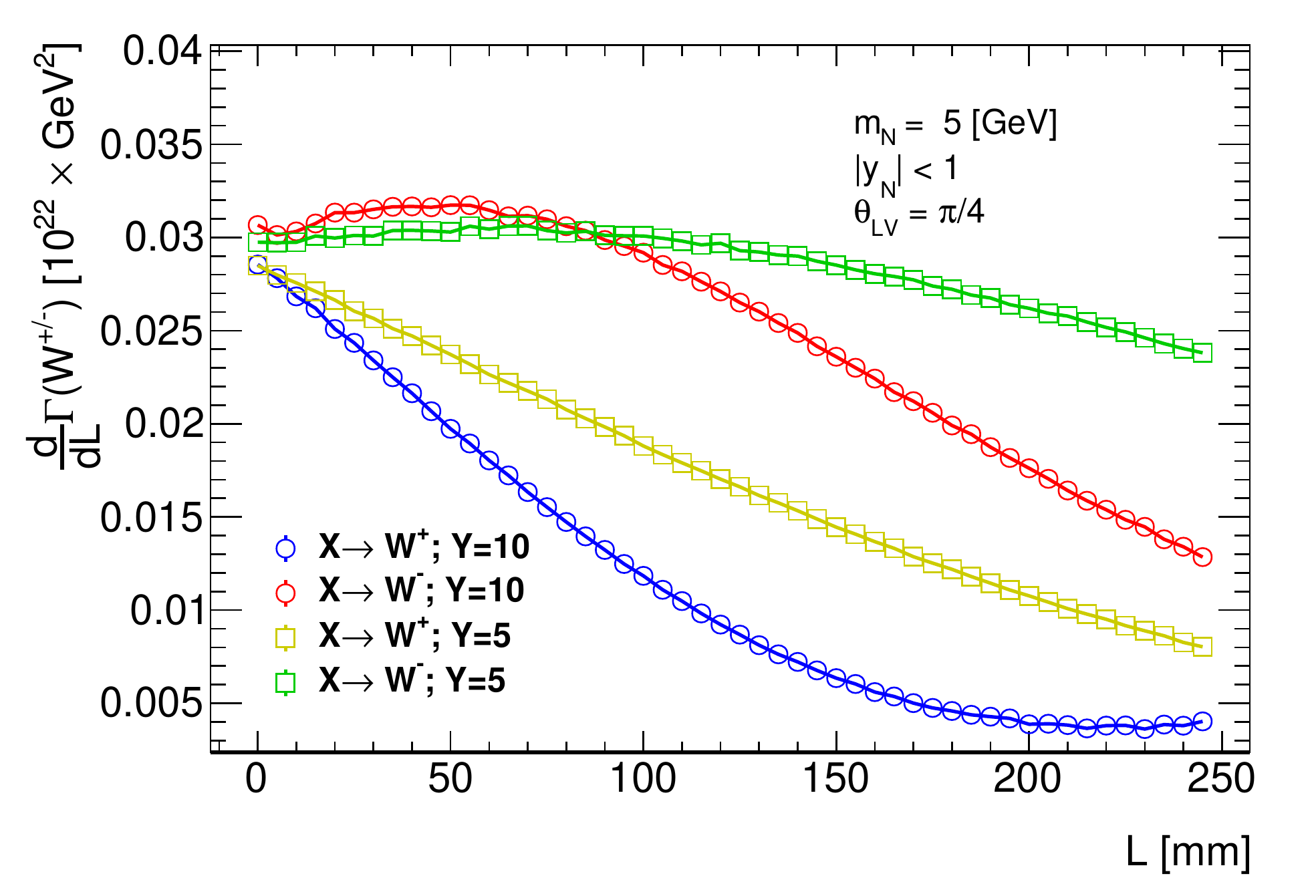}
\caption{Heavy neutrino oscillation modulation. Left panel: $M_N=5$ GeV, $\theta_{LV}=\pi/2$, $|y_{N}| < 1$ and $Y=5,10$. Right panel: $M_N=5$ GeV, $\theta_{LV}=\pi/4$, $|y_{N}| < 1$ and $Y=5,10$. }
\label{fig:oscparam1}
\end{figure}

\begin{figure}
\centering
\includegraphics[width=0.49\textwidth]{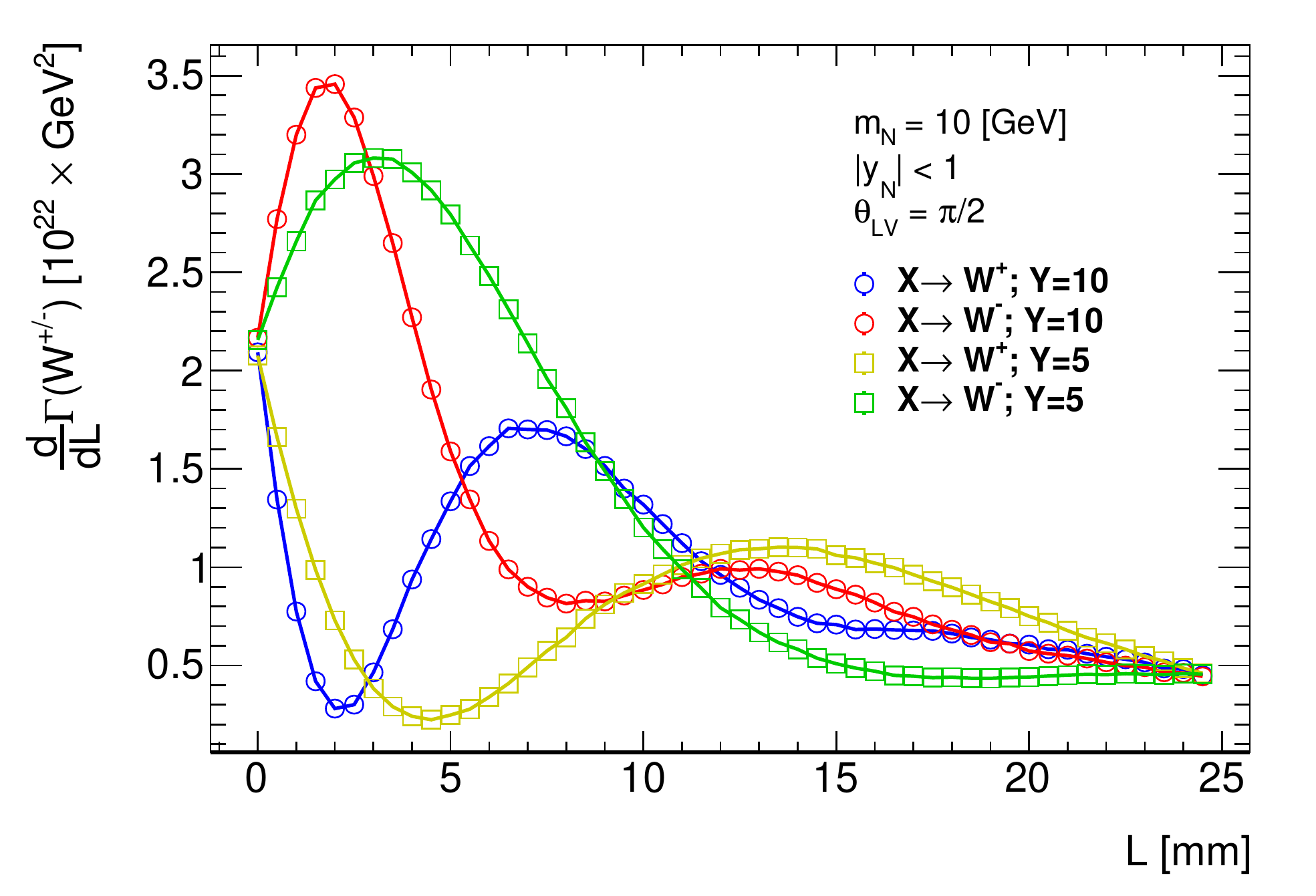}\hspace{0.01 cm}
\includegraphics[width=0.49\textwidth]{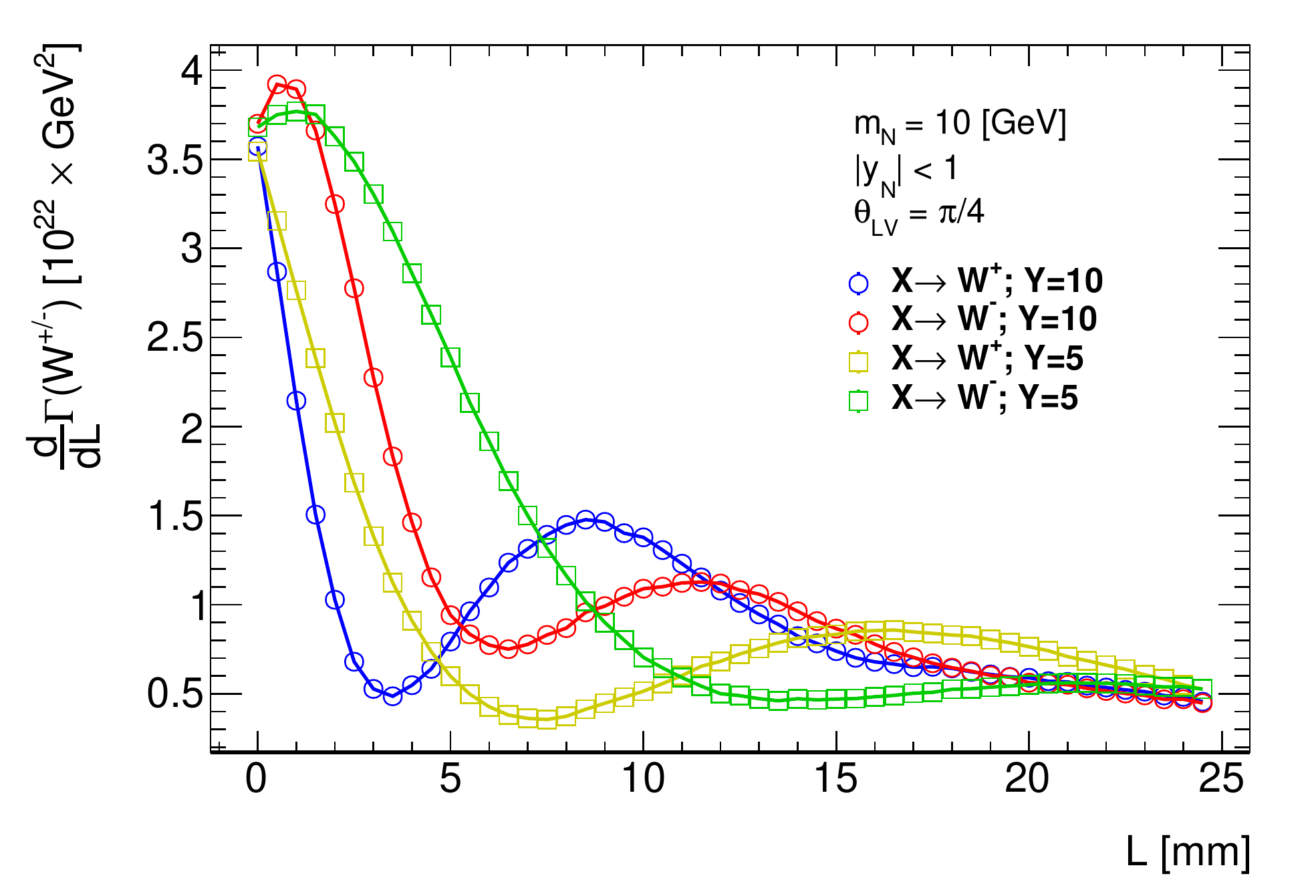}
\caption{Heavy neutrino oscillation modulation. Left panel: $M_N=10$ GeV, $\theta_{LV}=\pi/2$, $|y_{N}| < 1$ and $Y=5,10$. Right panel: $M_N=10$ GeV, $\theta_{LV}=\pi/4$, $|y_{N}| < 1$ and $Y=5,10$. }
\label{fig:oscparam2}
\end{figure}

It is important to point out that for masses $M_N>15$ GeV the heavy neutrinos $(N)$ tend to decay in a very short distance $\lesssim 1$ mm 
(Fig.~\ref{fig:lambda} right-panel), not offering good chances to observe the modulation of heavy neutrino oscillations. This can be seen also by considering the oscillation length $L_{\rm osc} = (2 \pi \gamma_N \beta_N /Y) (1/\Gamma_N )$ which is proportional to the inverse decay width $1/Γ_N \propto 1/M_N^5$, i.e., $L_{\rm osc}$ decreases fast when $M_N$ grows. Furthermore, when $M_N$ increases, the differential decay width has stronger exponential attenuation and, consequently, the oscillation effects are more difficult to detect. On the other hand, 
for masses $M_N<5$ GeV (Fig.~\ref{fig:lambda} left-panel) ${\gamma_N \beta_N}$ is very large, the factor $1/({\gamma_N \beta_N})$ appearing in 
the exponential in (Eqs.~\ref{effdwfosc} and \ref{effdwfosc2}) is small and strongly suppresses the considered modulation quantity $d \Gamma/d L$.
Therefore, in order to select the events for feasible measurement of the modulation of $d \Gamma/d L$ at the LHC detectors \cite{Aad:2008zzm,Chatrchyan:2008aa},
 we require that the rapidity of the heavy neutrino $(N)$ satisfies $|y_{N}| < 1$, 
 i.e., $\gamma_N \beta_N$ is small\footnote{Rapidity $y_N$ is defined as: $\gamma_N \beta_N = \sinh y_N$. Events
 with $|y_{N}| \gg 1$ are not realistic because the acceptance of detectors has a limit at $|y_{N}| \approx 2$.}. 
 The quantity $\gamma_{N}\beta_{N} = |{\vec p}_N({\hat p}'_N)|/M_N$ is now treated as a random variable which is used to 
 re-evaluate Eq.~\ref{effdwfosc} several times (i.e., Eq.~\ref{effdwfosc2}) in steps of $L$ and for different
 choices of $M_{N}$ and $\theta_{LV}$. The re-evaluation is done 10,000 times in each step of $L$, for fixed $M_{N}$ and $\theta_{LV}$, the average of all 
 those values is used as the new expected value for $d \Gamma/d L$. The phase-space where the measurement can be performed and the resolution of the 
 detector (see Ref.~\cite{ATL-PHYS-PUB-2019-013}) are taken into account during the re-evaluation process. The comparison
 between $d\Gamma/dL$ when we used a fixed (and average) value of $\gamma_{N}\beta_{N}$, and $d\Gamma/dL$ re-evaluated when using the random 
 sampling of $\gamma_{N}\beta_{N}$ from the simulation as aforementioned is shown in Fig.~\ref{fig:fixvarcom}.
Figs.~\ref{fig:oscparam1} and \ref{fig:oscparam2} show the results of evaluating Eq.~\ref{effdwfosc} with variable values of $\gamma_N \beta_N$, as previously explained, 
for different choices of $M_N$, $\theta_{LV}$ and $Y(\equiv \Delta M_N/\Gamma_N)$. We point out that the distribution of $\gamma_N \beta_N$ associated 
with $W^+$ decays is not the same as the one obtained with $W^-$ decays, despite the severe cut $|y_N|<1$ that we apply (cf.~also Fig.~\ref{fig:lambda} left 
panel for the full averaged values); this fact causes a significant difference between the results with $W^+$ and those with $W^-$.


According to previous calculations of the expected cross section, and taking into account the luminosity collected at the LHC during the Run-II and the luminosity 
expected for the future High Luminosity LHC (HL-LHC) \cite{Apollinari:2017cqg}, the corresponding number of events could be in tens of thousands  
\cite{Drewes:2019fou,Liu:2019ayx}. We simulate the considered process for a benchmark of 100 and 1000 observed events, considering a 
detector resolution of the position of the secondary vertex equal to 0.3 mm \cite{ATL-PHYS-PUB-2019-013} modeled with a gaussian distribution.
\begin{figure}
\centering
\includegraphics[width=0.49\textwidth]{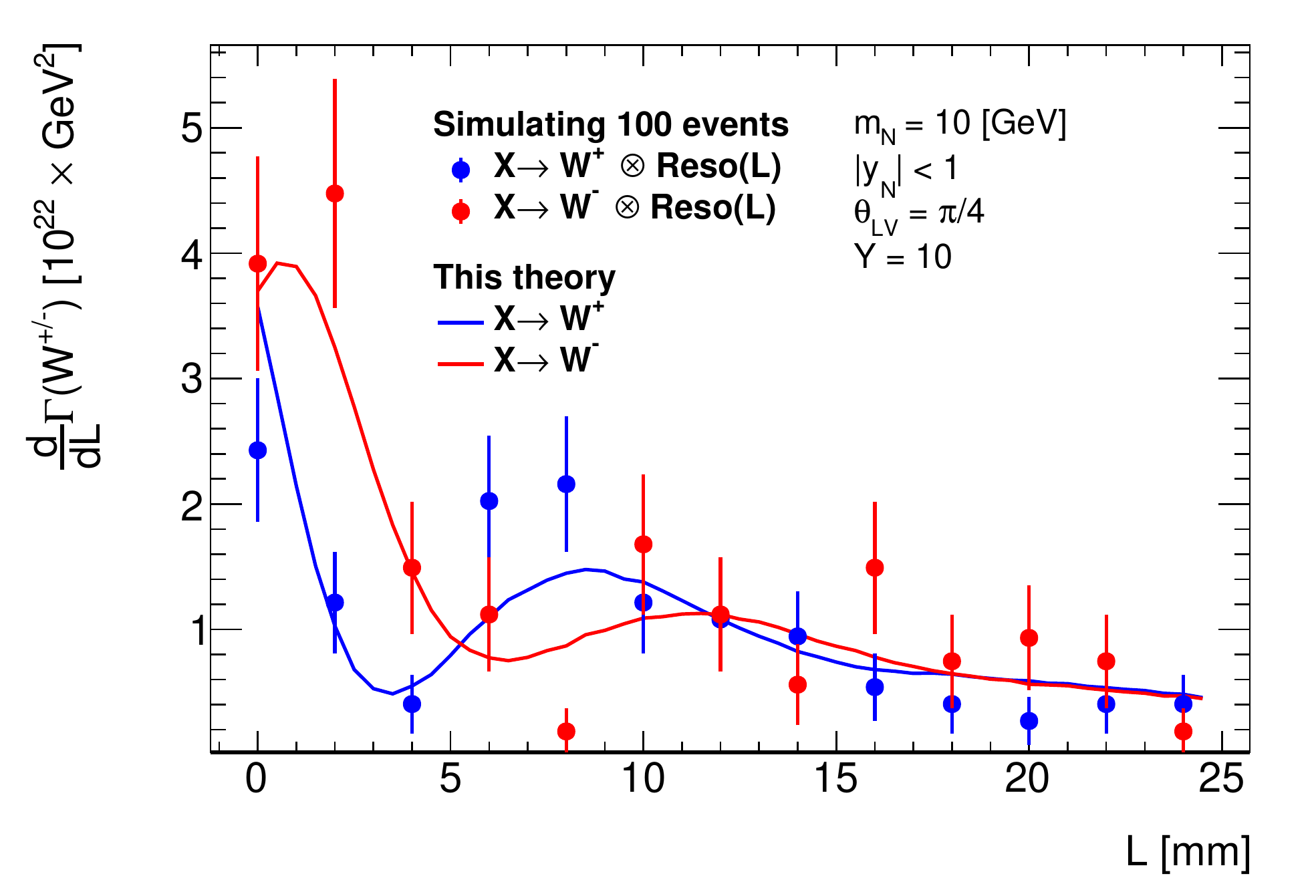}\hspace{0.01 cm}
\includegraphics[width=0.49\textwidth]{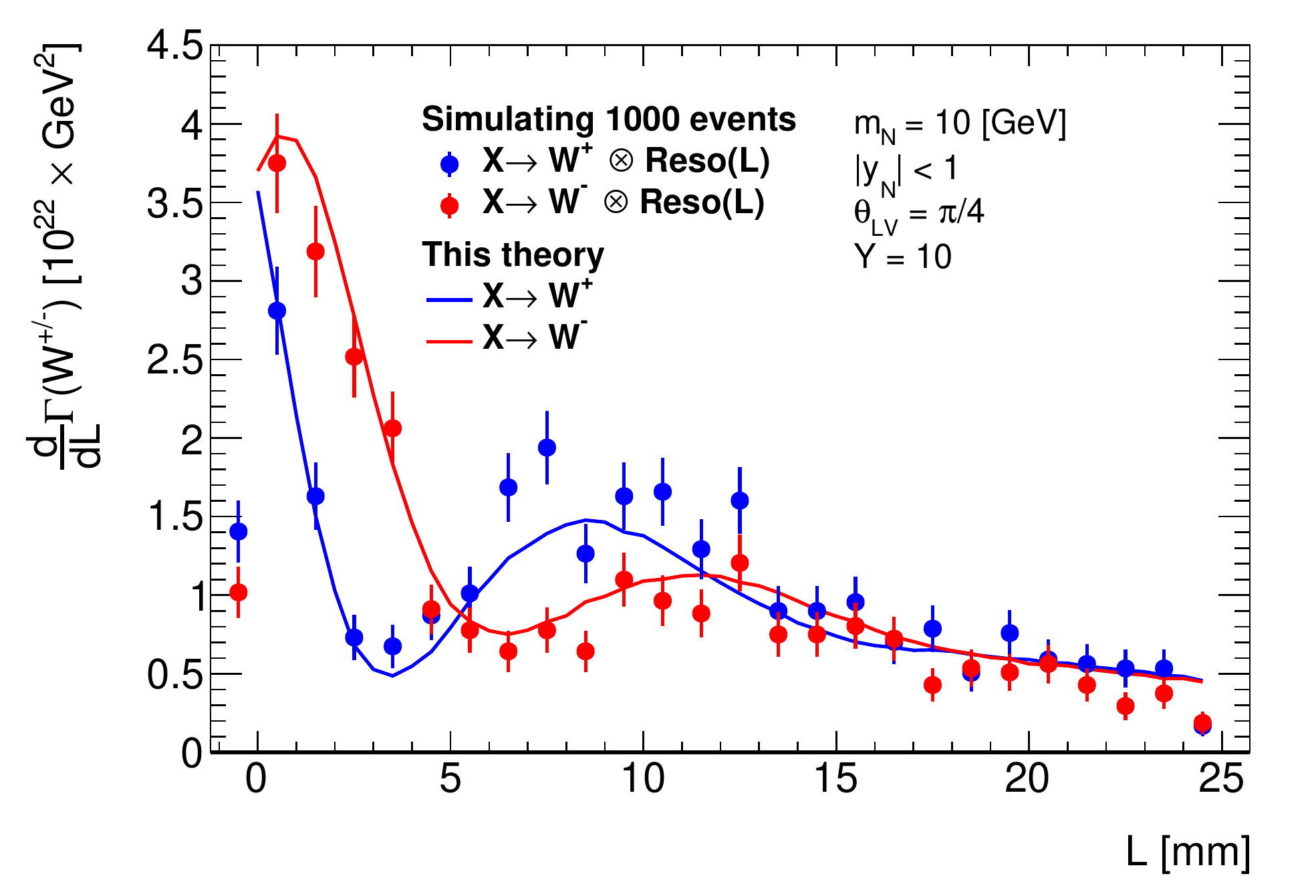}
\caption{The solid line corresponds to the heavy neutrino oscillation modulation, and the data-points correspond to random samplings of the heavy neutrino oscillation modulation convolved with the detector resolution. Left panel: $M_N=10$ GeV, $\theta_{LV}=\pi/4$, $|y_{N}| < 1$, $Y=10$ and 100 simulated events. Right panel: $M_N=10$ GeV, $\theta_{LV}=\pi/4$, $|y_{N}| < 1$, $Y=10$ for 1000 simulated events.}
\label{fig:oscparam3}
\end{figure}
We point out that the scenario considered in Fig.~\ref{fig:oscparam3} represents the worst possible for the measurement of the modulation and the CP phase $\theta_{LV}$.
If the number of events in the simulation is further increased, and the resolution of the vertex is taken to be zero~mm, the results of Fig.~\ref{fig:oscparam3} 
(right panel) will coincide with the curves (``This theory'') and with the corresponding results of Fig.~\ref{fig:oscparam2} (right panel).

Similar results are obtained for the oscillation effects in the lepton number conserving (LNC) case ($\mu^{\pm} \tau^{\mp}$) with the 
replacements $\theta_{LV} \mapsto \theta_{LC}$, cf.~Ref.~\cite{Cvetic:2015ura}.  In the case of Dirac neutrino $(N)$ the total decay width $\Gamma_N$ has by about $40\%$  
lower values of the mixing coefficients ${\cal N}_{\ell N}$ in Eq.~\ref{DNwidth} (cf.~Fig.~2 in Ref.~\cite{Cvetic:2015naa}). LNC type of rare processes may be more difficult
to identify experimentally due to a possibility of the larger backgrounds.
\section{Discussion of the results and summary}
\label{sec:dis}

In this work we have considered the oscillation modulation of $d \Gamma/d L$, for the LNV process \\
$W^{\pm} \to \mu^{\pm} N \to \mu^{\pm} \tau^{\pm} W^{\mp *}$ $\to \mu^{\pm} \tau^{\pm} e^{\mp} \nu_e$ at the LHC, 
in the scenario of two almost degenerate (on-shell) Majorana neutrinos $(N_j)$. We found out that, for the measurement of the modulation of $d \Gamma/d L$, the heavy neutrino mass $M_N$ should be neither 
very high ($> 15$ GeV) nor low ($< 5$ GeV). This is so because, for our purposes, the heavy neutrinos $(N)$ should neither decay at a too short distance ($ \lesssim 1$ mm) nor should the exponential 
factor in $d \Gamma/d L$ lead to a too strong suppression of this quantity. 

According to Ref. \cite{Drewes:2019fou} and References therein there are scenarios for our type of process where tens of thousands of events with displaced vertices could be observed at LHC, after an appropriate background removal analysis. However, in our analysis we adopted a more conservative attitude, by assuming observation of 100 to 1000 events. We mention that there are several processes in the standard model that can produce a displaced vertex with an attached lepton, like Kaons decay-in-flight and b-hadrons semi-leptonic decays. Those processes have been extensively studied in ATLAS and CMS \cite{Aad:2019kiz,CMS:2014wda} , as backgrounds for long-lived neutral particle searches. As it is known, the distribution of the decay length of all these standard model processes is monotonically (exponential-like), decreasing as shown in figure 4 (right) of Ref. \cite{CMS:2014wda}.  Therefore, none on the studied processes are comparable to the background distribution shape.

In addition, we have observed that the simulation of the production of on-shell heavy neutrinos $N$ in LHC gave a different distribution of the values of the heavy neutrino kinematic quantity $\gamma_N \beta_N$ in 
the case of $W^+$ and $W^-$, because of different 3-momenta distributions of the produced $W^+$ and $W^-$. As a consequence, also the averages $\langle \gamma_N \beta_N \rangle$ are 
different in the two cases (cf.~Fig.~\ref{fig:lambda}). In comparison with our previous work  \cite{Cvetic:2018elt} where $\gamma_N \beta_N$ value was fixed ($\gamma_N \beta_N = 2$ ), here the mentioned distribution of $\gamma_N \beta_N$ was taken into account, and in addition we used the rapidity cut $y_N < 1$ ($\gamma_N \beta_N \equiv {\rm sinh}  y_N$) to avoid a too strong suppression of $d \Gamma/d L$. As a consequence, we found out that the modulation is significantly smeared  by the fact that we have a distribution of (small) values  of $\gamma_N \beta_N$ and not a fixed (average) value (cf.~Fig.~\ref{fig:fixvarcom}).

We have also calculated the behavior of $d \Gamma/d L$ for $W^+$ and $W^-$ cases and for various values of the parameters: $M_N=5$ GeV and $10$ GeV; $Y \equiv \Delta M_N/\Gamma_N=5$ and $10$; and the CP 
phase $\theta_{LV}=\pi/2, \pi/4$. The number of events was assumed to be (almost) infinite and the vertex resolution was considered ideal ($0$ mm), cf.~Figs.~\ref{fig:oscparam1} and \ref{fig:oscparam2}. The form of the 
modulation of $d \Gamma/d L$ turned out to have a strong dependence on the value of the CP-phase  $\theta_{LV}$, which indicates a possibility to extract the value of $\theta_{LV}$ from measurements of such modulations.
The dependence on the parameter $Y \equiv \Delta M_N/\Gamma_N$ was also significant. When $M_N=5$ GeV and $\theta_{LV}=\pi/2$, we observed from Fig.~\ref{fig:oscparam1} (left panel) that 
for $W^+$ decays inside the region $0 \leq L \leq 190$ mm the number of expected events is bigger for $Y=5$ than $Y=10$; and inside $190 \leq L \leq 250$ mm the opposite is true. 
On the other hand, when $M_N=5$ GeV and $\theta_{LV}=\pi/4$, we observed from Fig.~\ref{fig:oscparam1} (right panel) that for $W^+$ decays inside the entire region $0 \leq L \leq 250$ mm 
the number of expected events is bigger for $Y=5$ than $Y=10$. For the $W^-$ decays the comparisons change significantly: when $M_N=5$ GeV and $\theta_{LV}=\pi/2$, for $W^-$ decays inside the 
region $0 \leq L \leq 190$ mm the number of expected events is bigger for $Y=10$ than $Y=5$; and inside $190 \leq L \leq 250$ mm the opposite is true; when $M_N=5$ GeV and $\theta_{LV}=\pi/4$, only 
inside the region $0 \leq L \leq 90$ mm is the number of expected event bigger for $Y=10$ than $Y=5$, and for $90 \leq L \leq 250$ mm the opposite is true. When the heavy neutrino mass is higher, $M_N=10$ GeV, 
it turned out that the modulation of $d \Gamma/d L$ practically vanishes for $L>20$ mm, cf.~Fig.~\ref{fig:oscparam2} where left (right) panel represents $\theta_{\rm LV} =\frac{\pi}{2}~(\frac{\pi}{4})$; on the other hand, for $L < 10$ mm the modulations turned out to be strong.

In conclusion, in Fig.~\ref{fig:oscparam3} the results were presented (at $M_N=10$ GeV; $Y=10$ and $\theta_{LV}=\pi/4$) for a more realistic case, i.e., when the total number of the detected  LNV
events is finite, either 100 (left panel) or 1000 (right panel). In addition, the resolution of the secondary vertex position was taken to be nonzero, e.~g.,  $0.3$ mm. In the case of 100 simulated events, the modulation with enough 
statistical significance was observed for distances up to $L=6$ mm; whereas in the case of 1000 simulated events, the observable modulation increased up to $L=12$ mm.

Finally, in this work we considered the scenario of two heavy almost degenerate neutrinos $(N_j)$ with masses within $1$ GeV $\leq M_N \leq 10$ GeV. We have evaluated the possibility to measure an 
LNV oscillation process in such a scenario where the modulation of the quantity $d \Gamma/d L$ for the process 
$W^{\pm} \to \mu^{\pm} N \to \mu^{\pm} \tau^{\pm} W^{\mp *}$ $ \to \mu^{\pm} \tau^{\pm} e^{\mp} \nu_e$ at the LHC can be measured within the detector. Here $L$ is the distance (within the detector) between the two vertices 
of the process. We have found out some realistic conditions where 
$|B_{\mu N}|^2 \sim |B_{\tau N}|^2 \sim 10^{-6}$, $M_N \approx 5$ GeV$-10$ GeV and $Y (\equiv \Delta M_N/\Gamma_N) \sim 10$. To measure such a process, we also pointed 
out the importance of the application of the rapidity cuts, $y_N < 1$.

\section{Acknowledgments}
This work is supported in part by FONDECYT Grant No.~3180032 (J.Z.S.) and FONDECYT Grant No.~1180344 (G.C.). The work of S.T.A. is supported by the National Science Foundation (NSF) grant 1812377.
The work of A.D. is  supported by the Japan Society for the Promotion of Science (JSPS) Postdoctoral Fellowship for Research in Japan. 

\appendix


\end{document}